\documentclass{article}

\usepackage{graphicx}        
\usepackage{multicol}        
\usepackage[bottom]{footmisc}
\usepackage{url}   
\usepackage{subfigure}   

\usepackage[colorlinks=true,linkcolor=black,anchorcolor=black,citecolor=blue,filecolor=red,menucolor=black,urlcolor=blue,breaklinks=true,pdfhighlight=/P,pdfmenubar=true,pdftoolbar=true,pdfpagelabels=true,pdfstartpage=1,pdfstartview=FitV,pdftitle={Quickest Paths in Simulations of Pedestrians},pdfsubject={Pedestrian Route Choice},pdfauthor={Kretz},pdfcreator={Kretz},pdfproducer={Kretz},pdfkeywords={pedestrian route choice simulation VISSIM assignment quickest path space time telepresence behavior}]{hyperref}	
\usepackage[numbers,sort&compress]{natbib}	
\usepackage{hypernat} 

\usepackage{simplemargins}
\setallmargins{1in}

\begin{document}

\title{Using a Telepresence System to Investigate Route Choice Behavior}
\author{Tobias Kretz$^1$, Stefan Hengst$^1$,  Antonia P\'{e}rez Arias$^2$, \\ Simon Friedberger$^2$, and Uwe D. Hanebeck$^2$ \\ \\
$^1$: PTV Planung Transport Verkehr AG \\ Stumpfstr. 1, D-76131 Karlsruhe \\ 
\texttt{first.last@ptv.de}\\ \\
$^2$: Intelligent Sensor-Actuator-Systems Laboratory (ISAS). \\ Karlsruhe Institute of Technology (KIT). Germany. \\ \texttt{first.last@kit.edu}}
%
%
\maketitle

\begin{abstract}
A combination of a telepresence system and a microscopic traffic simulator is introduced. It is evaluated using a hotel evacuation scenario. Four different kinds of supporting information are compared, standard exit signs, floor plans with indicated exit routes, guiding lines on the floor and simulated agents leading the way. The results indicate that guiding lines are the most efficient way to support an evacuation but the natural behavior of following others comes very close. On another level the results are consistent with previously performed real and virtual experiments and validate the use of a telepresence system in evacuation studies. It is shown that using a microscopic traffic simulator extends the possibilities for evaluation, e.g. by adding simulated humans to the environment.
\end{abstract}

\section{Introduction}

To accurately plan and evaluate safety measures in buildings and transportation systems, knowledge of the behavior of pedestrians in emergency situations is required. There are two major ways to acquire this information, conducting experiments or performing computer simulations. The problem with experiments is the immense effort that is necessary to recreate an appropriate situation. This is even more difficult in the planning phase when a prototype would have to be constructed. In addition many situations cannot be evaluated experimentally because the safety of the test subjects cannot be guaranteed. Simulations on the other hand are usually based on relatively simple models of pedestrian behavior and cannot capture the complexities of human thought processes and decision making. A common solution is to conduct simple experiments and simulations and attempt to derive generally applicable rules from them \cite{Chattaraj2009}. The generality makes these rules necessarily crude; an example would be the required exit width depending on the number of people in a building.
Kobes et al. \cite{Kobes2009} propose the usage of so called serious games to combine the benefits of simulations and experiments. Their system allows a test subject to try to escape from a simulated emergency situation using common video game techniques. The problem with this approach is that the user doesn't actually move even though proprioception (i.e. self-perception of motion) has been shown to improve navigational abilities. We therefore introduce a combination of an extended range telepresence system and a pedestrian simulation.
The effectiveness of the system is shown by comparing different utilities for finding fire exits. Such different utilities, like escape exit signs or guiding lines, can be seen and followed with varying simplicity. As for example guiding lines are visually very intrusive and may be considered unaesthetic a quantitative evaluation of the usefulness of different signage is desirable.

\section{Combined System}

This section presents the components of our experimental setup; the extended range telepresence system and the pedestrian simulation software.

\subsection{Extended Range Telepresence}
The extended range telepresence system allows a user to feel present in a remote or virtual environment (called the target environment) by locally reproducing perception for the user and remotely reproducing actions by the user. To achieve this, the user is wearing a head-mounted display (HMD) capable of displaying the target environment in 3D and playing back sound (cf. Figure 1). The head-mounted display is fitted with additional sensors that allow its position and orientation in the room to be tracked. When the user takes a step forward, this movement will be registered and transferred to the target environment and what the user sees will change accordingly. This process is further complicated  by an algorithm called Motion Compression \cite{Nitzsche2002,Nitzsche2004,Roessler2004}, which allows the target environment to be much larger than the user environment. The path of the user is curved to require less space while keeping length and turning angles of the paths in both environments identical. It has been shown that users do not notice slight changes in curvature \cite{Nitzsche2003}. The image the user sees is then rotated slightly to steer him on the calculated user path.

\subsection{VISSIM}
We have connected the pedestrian and vehicle simulation software VISSIM \cite{Fellendorf2010,VISSIM2010} to the extended range telepresence allowing us to simulate environments that include virtual agents (cf. Figure 2). These virtual (simulated) agents react to the telepresent user as if he was a simulated agent allowing him to become part of the simulation and interact with it.

\begin{figure}[htbp]
\begin{center}
\includegraphics[width=0.500\columnwidth]{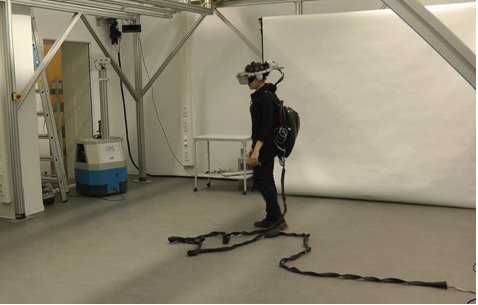}
\caption{User in the telepresence System, wearing HMD and backpack computer for processing.}
\label{fig:1}
\end{center}
\end{figure}
  
\begin{figure}[htbp]
\begin{center}
\includegraphics[width=0.500\columnwidth]{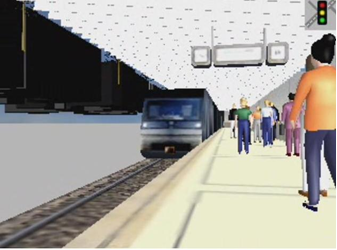}
\caption{Screenshot from the VISSIM microscopic traffic simulator.}
\label{fig:2}
\end{center}
\end{figure} 

\subsection{Benefits}
This setup has several benefits apart from proprioception. Three dimensional vision with the head mounted display allows to naturally judge distances and increases realism. The user position is tracked, which allows the simple creation of detailed records of his movement. The orientation of the head-mounted display is tracked as well which can be used to extract coarse focus of attention information, e.g., whether a fire exit is visible. Using a pedestrian simulation enables us to populate the simulated world with other humans without requiring further test subjects. This allows conducting experiments concerning the behavior of individuals in large crowds quickly and cheaply.

\section{Route Choice behavior In a Hotel Evacuation}

\subsection{Scenario}
\begin{figure}[htbp]
\begin{center}
\includegraphics[width=0.500\columnwidth]{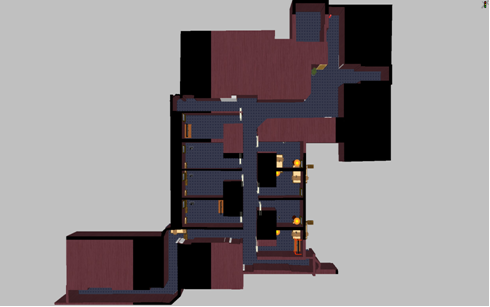}
\caption{Orthogonal view of the 3D scenario.}
\label{fig:3}
\end{center}
\end{figure}

The combined VISSIM-Telepresence system is used to study the influence of different signage for finding fire exits in a virtual hotel scenario (Fig. \ref{fig:3}), it has the same layout as the scenario used by Kobes et al. in \cite{Kobes2007fire,Kobes2009,Kobes2010case,Kobes2010way}. The layout was enhanced with textures and furniture in order to realistically reproduce a typical hotel scenario.

Using this scenario has two advantages: First, the layout of the chosen hotel is classified as complex \cite{Kobes2010case} and the choice of the nearest exit is not trivial. Second, this scenario reproduces a real hotel and data of a real case study performed in the hotel are available in \cite{Kobes2010case}, so that we can compare our results with the real data.

\subsection{Experiment Description}
\begin{figure}[htbp]
\begin{center}
\includegraphics[width=0.500\columnwidth]{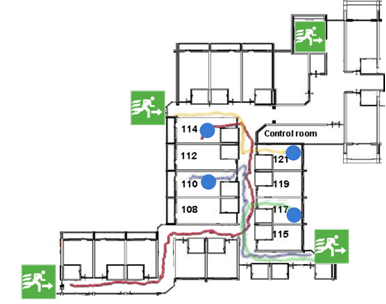}
\caption{Example trajectories. Base picture adapted from \cite{Kobes2007fire}.}
\label{fig:4}
\end{center}
\end{figure}

\begin{figure}[htbp]
\begin{center}
\includegraphics[width=0.500\columnwidth]{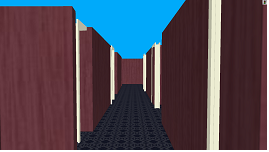}
\caption{Screenshot of evacuation scenario without supporting information.}
\label{fig:5}
\end{center}
\end{figure}

Preliminary experiments were conducted to check the user behavior concerning the virtual walls. In principle, the user could consciously decide to move through the virtual walls. Different to an experiment designed to be done in front of a screen and controlled by keys, mouse or joystick, the telepresence system has no mechanism to prevent this. The preliminary experiments showed that the visual information that the user receives via the head-mounted display is sufficient to navigate in the virtual environment without colliding with the virtual walls (Fig. \ref{fig:4} shows exemplary trajectories of these experiments). However, without supporting information (cf. Fig. \ref{fig:5}) most test subjects were unable to find the nearest exit in these experiments.

\begin{figure}[htbp]
\begin{center}
\includegraphics[height=75pt]{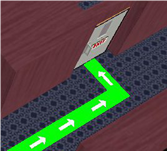} \hspace{6pt}
\includegraphics[height=75pt]{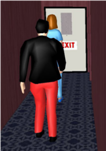} \hspace{6pt}
\includegraphics[height=75pt]{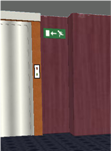} \hspace{6pt}
\includegraphics[height=75pt]{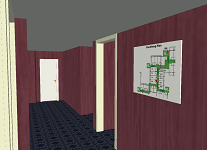}
\caption{Screenshots of scenarios with supporting information: (a) guiding lines, (b) simulated agents, (c) escape exit signs, and (d) floor plan.}
\label{fig:6}
\end{center}
\end{figure}

A case study was designed to investigate the influence of the signage in finding the nearest exit in case of a hotel evacuation. As part of this study, we compared exit choice, travel times, walking distances, and walking speeds towards exits under following conditions:

There is a guiding line on the floor (Fig. \ref{fig:6}(a)).
There are other persons (simulated agents) walking to the exit (Fig. \ref{fig:6}(b)).
There is standard escape exit signage above head (Fig. \ref{fig:6}(c)).
There is an evacuation floor plan in the room where the evacuation starts (Fig. \ref{fig:6}(d)).

\subsection{Experiment Participants}
We introduced 20 participants, all male, between 21 and 32 years old to the scenario. The participants had the opportunity to familiarize themselves with the telepresence system in a simpler scenario. All participants tested the 4 conditions. The participants started each test run at different positions to avoid learning effects. Moreover, 5 participants started with condition 1, 5 with condition 2, etc, in a way that we could analyze the performance under different signage conditions with and without learning effects separately. The participants were instructed to leave the building as fast as possible as they would be a real evacuation. 

\subsection{Performance Measures}
In order to evaluate the efficiency of the supporting information, exit choice, travel times, walking distances, and walking speeds towards exits were recorded. In order to quantify the subjective preference of the participants for each type of signal, a questionnaire was used.

\section{Results}

The participants were not able to find the nearest exit without supporting information. Therefore, the next results only report the performance measures for the scenarios with supporting information.

\subsection{Objective Measures}

\subsubsection{Nearest Exit Choice}
\begin{figure}[htbp]
\begin{center}
\includegraphics[width=0.500\columnwidth]{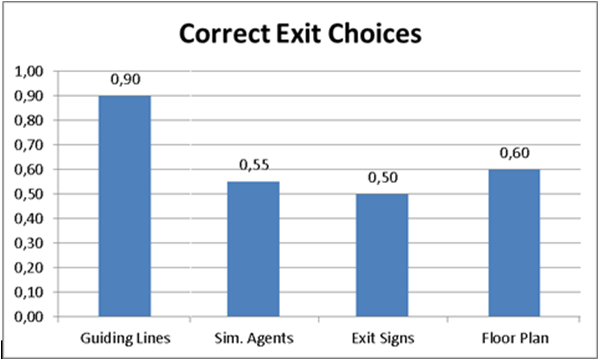}
\caption{Percentage of correct exit choice for each signage condition.}
\label{fig:7}
\end{center}
\end{figure}

The evaluation of the exit choice in Fig. \ref{fig:7} shows that the guiding lines are the most efficient signage to find the nearest exit. By using other signage the nearest exit is only chosen in about 50\% of the test runs.
 
\subsubsection{Travel Time}
\begin{figure}[htbp]
\begin{center}
\subfigure[ ]{\includegraphics[width=0.450\columnwidth]{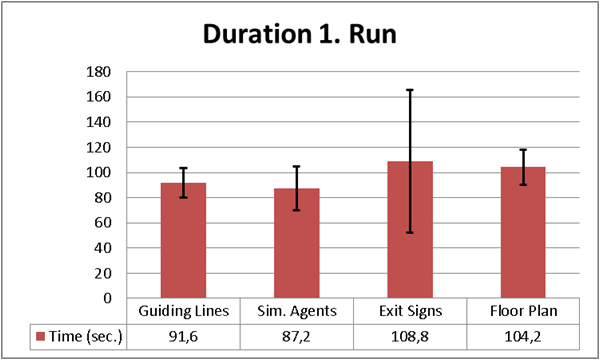} \hspace{6pt}
\label{fig:9a}
}
\subfigure[ ]{\includegraphics[width=0.450\columnwidth]{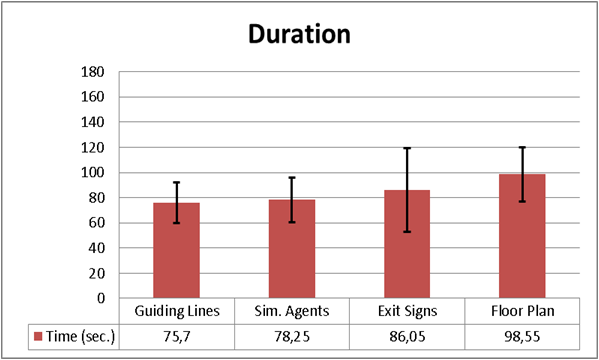}
\label{fig:9b}
}
\caption{Average duration for each signage condition using only the first runs of the participants \subref{fig:9a}. Average duration using all runs of the participants \subref{fig:9b}.}
\end{center}
\end{figure}

The time needed to find the exit in the evacuation scenario is shown in Fig.~\ref{fig:9a}. When only the first test runs are evaluated (i.e., the participants do not know the hotel scenario in advance), the guiding lines and the presence of other pedestrians lead to the shortest travel times with average values of 91.6 sec. and 87.2 sec., respectively. Escape exit signs and the escape floor plan have the longest travel times. Note that there is a higher standard deviation across the participants when using the exit signs, whereas the floor plan leads to longer times for most participants.


The guiding lines and the presence of other pedestrians also lead to faster evacuations when regarding the average time duration for all test runs Fig.~\ref{fig:9b}, although the duration of the evacuation using the exit signs and the floor plan is as expected shorter when the user knows the building in advance.


\subsubsection{Walking Distance}
\begin{figure}[htbp]
\begin{center}
\subfigure[ ]{\includegraphics[width=0.450\columnwidth]{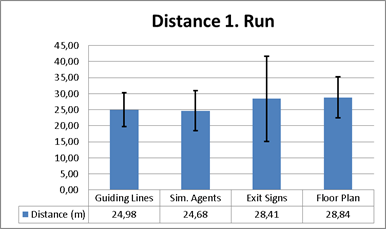} 
\label{fig:10a}}
\subfigure[ ]{\includegraphics[width=0.475\columnwidth]{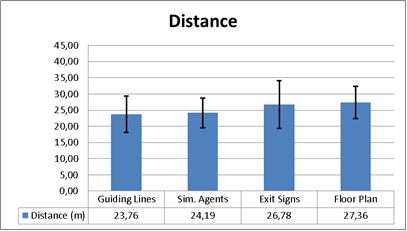}
\label{fig:10b}}
\caption{Average walking distance using only the first runs of the participants \subref{fig:10a}. Average walking distance using all runs of the participants \subref{fig:10b}.}
\end{center}
\end{figure}

Fig.~\ref{fig:10a} shows the average of the covered distances to the exits using only the first runs of the participants. The guiding lines and the presence of other pedestrians again lead to shorter walking distances than the exit signs and the floor plan. The same trend is observed when considering all test runs of the participants (Fig.~\ref{fig:10b}).

\subsubsection{Walking Speed}
\begin{figure}[htbp]
\begin{center}
\includegraphics[width=0.500\columnwidth]{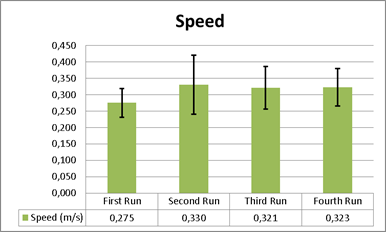}
\caption{Average velocity of participants at each run.}
\label{fig:12}
\end{center}
\end{figure}

The learning effect is clearly observed by regarding the average velocity of the participants at each test run in Fig. \ref{fig:12}. The average velocity in the first test run is significantly lower than in the other runs. However, no significant difference is observed between the second run and the next runs.

\subsection{Subjective Measures}
The evacuation scenario was found to be modeled realistically by almost all (19) participants. Most participants found the guiding lines to be the most efficient signal. Moreover, in case of fire, 19 participants would prefer the guiding lines and 7 participants would prefer a combination of guiding lines and exit signs (Table \ref{tab:1}).
\begin{table}[htbp]
\begin{center}
\begin{tabular}{|l|c|c|c|c|} \hline
	                                                &Guiding  	&Exit  	&Sim.  	&Floor  \\ 
	                                                & Lines	& Signs	& Agents	& Plan \\ \hline
Which aid seemed to provide the fastest way out?	&  9&	7&	2&	2\\ \hline
Which aid seemed the most useful?                 &	17&	2&	1&	0\\ \hline
Which aid would you prefer in case of fire?       &	19&	5&	2&	0\\ 
(Multiple answers possible)                       &   &  &   &   \\ \hline
\end{tabular}
\caption{Questionnaire used to evaluate the preference of the participants.}
\label{tab:1}
\end{center}
\end{table}

\subsection{Discussion}
In our route choice behavior study, the guiding lines turned out to be the most efficient signage method in order to find the nearest exit. This signage condition achieved shorter times and walking distances than exit signs or a floor plan hanging on the wall. The evaluation of the questionnaires also showed that participants have a clear preference for this signage condition, especially in the case of a fire evacuation. These results are in agreement with studies reported in \cite{Kobes2007fire, Proulx2000, Quellette1993} that indicate that photoluminescent low-level exit path markings are likely to be more effective compared to conventional escape route signs. 
The presence of other pedestrians turned out to be very beneficial for the evacuation in our study leading to exit times close to those of guiding lines. However, following other agents did not always guide to the nearest exit for the user who may have started in a different room.
In order to validate our extended range telepresence system as an adequate tool to perform such a route choice study, a comparison of our results with real data is necessary. For this purpose, we use the results from the case study in the hotel scenario presented in \cite{Kobes2010case}.
In our experiments not all participants chose the nearest exit, neither did they in the real experiments. The mean value of the covered distance to the chosen exit is 48.8 m in the real experiment, with a minimum value of 13.5 m and a maximum value of 83.2 m \cite{Kobes2010case}. The mean value of the covered distance in our experiments (considering only the first run) is 26.7 m, with minimum and maximum values of 13.5 m (following simulated pedestrians) and 54.5 m (following the exit signs), respectively. The mean and the maximum covered distance in extended range telepresence are rather lower than in the real experiments. However, all the values are within the range of distances achieved in the real experiments.
The mean value of the walking velocity in the real experiments \cite{Kobes2010case} is 1.03 m/s, which is higher than the mean velocity in our experiments. This difference may be due to the user being afraid of running outside the borders of the user environment or damaging the carried equipment. This difference is irrelevant for our evaluation of route choice as there are no differences in speed limiting factors along the different exit paths.

\section{Summary}

A combination of a telepresence system and a microscopic traffic simulator has been introduced and its efficacy for evaluating evacuation scenarios has been shown. As a first test scenario, the evacuation of a hotel using different kinds of signage has been evaluated. The results indicate that low-level exit path markings are the most efficient way of guiding people to an emergency exit but also that following others is efficient as well. These results are consistent with previously performed real and virtual experiments, which validates the use of our telepresence system in evacuation studies, and also shows the extended possibilities of using a pedestrian simulation software to add virtual agents.

\section{Acknowledgements}

This work was supported by the research project ``The Pedestrian Simulation VISSIM within a Telepresence System'' within the Central Innovation Programme for Small and Medium-sized Enterprises (ZIM) of the German Federal Ministry of Economics and Technology (BMWi). 

\nocite{_PED2008}

 \bibliographystyle{utphys2011b}
 \bibliography{telepresence-hotel}

\end{document}